%% file: main.tex
\documentclass[conference]{IEEEtran}
\IEEEoverridecommandlockouts
\usepackage{cite}
\usepackage{amsmath,amssymb,amsfonts}
\usepackage{algorithmic}
\usepackage{graphicx}
\usepackage{textcomp}
\usepackage{xcolor}
\usepackage{hyperref}
\usepackage{rotating}
\usepackage{multirow}
\usepackage{subcaption}
\usepackage{url}

\def\BibTeX{{\rm B\kern-.05em{\sc i\kern-.025em b}\kern-.08em
    T\kern-.1667em\lower.7ex\hbox{E}\kern-.125emX}}
\begin{document}

\hyphenation{Sci-BERT}

\title{GraphConfRec: A Graph Neural Network-Based Conference Recommender System\\
}

\author{\IEEEauthorblockN{Andreea Iana}
\IEEEauthorblockA{\textit{Data and Web Science Group} \\
\textit{University of Mannheim}\\
Mannheim, Germany \\
andreea@informatik.uni-mannheim.de}
\and
\IEEEauthorblockN{Heiko Paulheim}
\IEEEauthorblockA{\textit{Data and Web Science Group} \\
\textit{University of Mannheim}\\
Mannheim, Germany \\
heiko@informatik.uni-mannheim.de}

}

\maketitle

\begin{abstract}
In today's academic publishing model, especially in Computer Science, conferences commonly constitute the main platforms for releasing the latest peer-reviewed advancements in their respective fields. However, choosing a suitable academic venue for publishing one's research can represent a challenging task considering the plethora of available conferences, particularly for those at the start of their academic careers, or for those seeking to publish outside of their usual domain. 
In this paper, we propose GraphConfRec, a conference recommender system which combines SciGraph and graph neural networks, to infer suggestions based not only on title and abstract, but also on co-authorship and citation relationships. GraphConfRec achieves a recall@10 of up to 0.580 and a MAP of up to 0.336 with a graph attention network-based recommendation model. 
A user study with 25 subjects supports the positive results.

\end{abstract}


\section{Introduction}

Bibliographic data constitutes the third largest domain in the Linked Open Data (LOD) cloud, after geographic and social web data, as stated by the 2014 report on the State of the LOD cloud, which records 138 such datasets (13\%) \cite{schmachtenberg_adoption_2014}. LOD cloud's most recent version reports 150 datasets in this category (12\%).\footnote{\url{https://lod-cloud.net/} (last accessed 10/01/2020)}


SciGraph\footnote{\url{https://scigraph.springernature.com/explorer} (last accessed 17/04/2020)}, Springer Nature's successor of Springer's LOD Conference Portal \cite{birukou_springer_2017} is a LOD bibliographic set, encompassing information about journals, books, and book chapters published by Springer Nature since the 19\textsuperscript{th} century. In addition to publications, it also comprises metadata about corresponding events, such as conferences and workshops, as well as information regarding associated persons and institutions. 

On the one hand, bibliographic datasets have been largely utilised for post-hoc studies, such as scientometrics, in quantitative analyses of publications, citations, or authors \cite{dimou_visualizing_2014,hu_linked-data-driven_2013}. On the other hand, recommender systems developed for scientific publications either focus on recommending research papers \cite{birukou_multi-agent_2006,li_conference_2018}, or on suggesting academic venues, utilising the title or abstract of a paper \cite{forrester_new_2017,iana_building_2019,medvet_publication_2014}. Only few models incorporate knowledge of co-authorship and citation relationships through hand-crafted features, requiring domain knowledge \cite{luong_publication_2012,rollins_manuscript_2017}.

In contrast, in this paper, we focus on the graph nature of scientific data (e.g. citation and co-authorship graphs), and on exploiting this using recent advancements in graph neural networks to provide users with recommendations for future academic venues. We propose GraphConfRec, a service which recommends conferences to submit a paper to, based not only on title and abstract, but also on co-authorship and citation relationships. The recommender utilizes SciGraph for information on past conferences and publications, and WikiCfP for data on upcoming events. GraphConfRec aims to aid researchers getting accustomed to a new domain, or those seeking to publish results outside their usual community, to navigate the plethora of available conferences.  

The rest of the paper is structured as follows. Section \ref{sec:related_work} provides an overview of related work. The dataset used and the proposed recommendation techniques are described in Sections \ref{sec:data} and \ref{sec:recommendation_techniques}, respectively. Section \ref{sec:evaluation} introduces the experimental setup and discusses the evaluation results. We conclude with a summary and discussion of open issues and future work in Sections \ref{sec:limitations_future} and \ref{sec:conclusion}.

\section{Related Work}
\label{sec:related_work}

Although recommender systems for scholarly data have been researched for nearly 20 years \cite{birukou_multi-agent_2006,janssen_uplib_2003}, the majority of these focus on recommending research papers \cite{beel_research-paper_2016}, or on detecting whether a paper is within the scope of a venue \cite{ghosal_deep_2019}. More recently, several recommender systems have been developed to suggest venues based generally on the textual content of a manuscript (e.g. title, abstract, keywords).
Instances of such systems include, but are not limited to, Springer Nature's Journal Suggester\footnote{\href{https://journalsuggester.springer.com}{https://journalsuggester.springer.com}}, Wiley's Journal Finder\footnote{\href{https://journalfinder.wiley.com/search?type=match}{https://journalfinder.wiley.com/search?type=match}}, Elsevier's Journal Finder\footnote{\href{https://journalfinder.elsevier.com}{https://journalfinder.elsevier.com}}\cite{kang_elsevier_2015}, Research Square’s Journal Guide\footnote{\href{https://www.journalguide.com}{https://www.journalguide.com}}, IEEE Publication Recommender\footnote{\href{http://publication-recommender.ieee.org/home}{http://publication-recommender.ieee.org/home}}, Jane\footnote{\href{https://jane.biosemantics.org/}{https://jane.biosemantics.org/
}}\cite{schuemie_jane_2008}, or techniques proposed by \cite{kobs_where_2020,iana_building_2019,medvet_publication_2014}. In contrast, models proposed by Pan et al. \cite{luong_publication_2012} or Rollins et al. \cite{rollins_manuscript_2017} utilise explicit feature engineering to include clues about citation or authors networks. Pradhan et al. \cite{pradhan_cnaver_2020} recommend venues by enhancing the representations of a paper's title and abstract with meta-path features extracted from citation and bibliographic networks. Küçüktunç et al. \cite{kucuktunc_theadvisor_2013} and Boukhris and Ayachi \cite{boukhris_novel_2014} generate recommendations without considering the content of a prospective manuscript by extending bibliographic data with citation relationships.

Furthermore, recommender systems can be improved by leveraging LOD. In particular, open datasets providing detailed descriptions of items can constitute a strong asset for content-based recommenders, which rely on the items' attributes for computing their similarities \cite{di_noia_linked_2012,heitmann_using_2010}. 
While many of these approaches employ handcrafted features \cite{ristoski_hybrid_2014}, graph embeddings trained on LOD, such as TransE \cite{bordes_semantic_2014} or RDF2vec \cite{ristoski_rdf2vec_2019}, have also been utilised for recommender systems, since they aim to encode similar entities closely in a latent vector space. Hence, items which are projected closely in the vector space would be used as candidates for recommendation. However, such shallow transductive embedding techniques are only applicable to static graphs and cannot be utilised in recommender systems built on evolving graphs, as they do not use node features and are unable to generate embeddings for nodes not seen during training, without additional optimization \cite{hamilton_inductive_2017}. The citation and co-authorship graphs utilized for conference recommendations evolve over time, as new publications act as unseen nodes added to the existing graph at test time. Such dynamic graphs require an inductive embedding method, capable of computing latent representations for new nodes. 

The particularities of graphs, comprised of unordered nodes with varied-sized neighbourhoods, lacking spatial orientation and having diverse structures, pose additional challenges for traditional machine-learning algorithms \cite{zhang_deep_2018,zhou_graph_2018}. Recently, graph neural networks (GNNs) have been extensively researched to extend the ability of deep neural networks to extract meaningful patterns from large-scale datasets by leveraging the input’s structure  to non-Euclidean domains, such as graphs, through a rich variety of approaches.
Although GNNs have already been considered in recommendation problems \cite{ying_graph_2018,wenqi_graph_2019}, they have not yet been explored, to the best of our knowledge, in the context of recommending academic venues.

\section{Data}
\label{sec:data}

GraphConfRec aims to recommend conferences for a prospective manuscript. To this end, we utilise SciGraph for training the recommender system. Additionally, data mined from WikiCfP and Google Scholar Metrics is used to provide optional information on future conferences (e.g. submission dates, description), and their rankings, in the prototype implemented to demonstrate GraphConRec's functioning.  

\subsection{SciGraph}
\label{subsec:scigraph} 

\begin{figure}[htbp]
\centering
\includegraphics[width=0.4\textwidth]{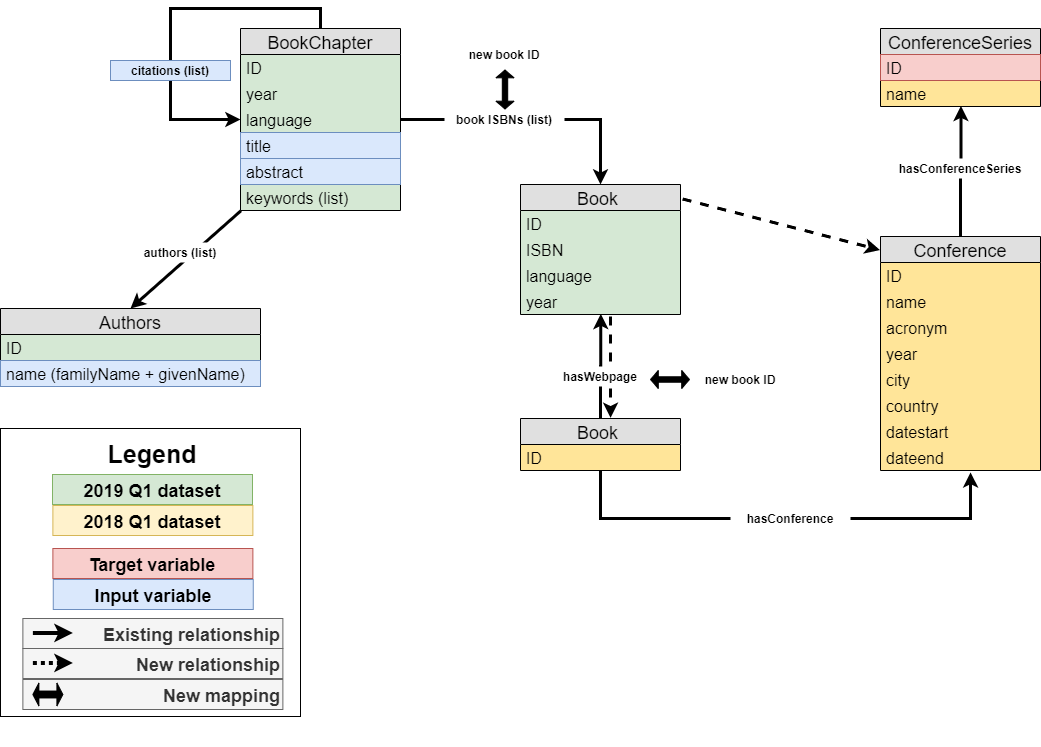}
\caption{Data model of SciGraph subset used in GraphConfRec.} 
\label{fig:SciGraph_datasets}
\end{figure}

SciGraph serves as the primary data source for training the recommender system. SciGraph's latest version contains information on approximately 4.5M book chapters, 270k books published by Springer Nature since 1839 and over 7M related persons. 
Only a fraction of the books contained in SciGraph correspond to conference proceedings. Moreover, the latest datasets do not encompass conference information. Therefore, we created a mapping between the latest 2019 Q1 SciGraph release, which incorporates citation information, and the 2018 Q1 release, which includes information about conferences, but not about referenced works. Fig. \ref{fig:SciGraph_datasets} illustrates the SciGraph subsets used for recommendation from these two datasets, namely information on abstracts, titles, citations, and authors (the main attributes are highlighted). 

\begin{figure}[htbp]
    \centering
    \begin{subfigure}[c]{0.22\textwidth}
        \includegraphics[width=\textwidth]{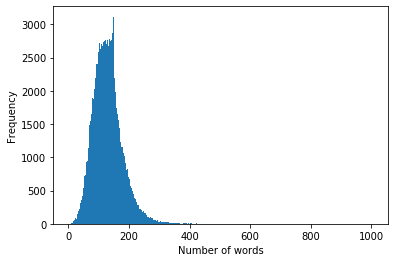}
        \subcaption{Average abstract length.}
        \label{subfig:words_per_abstract}
    \end{subfigure}
    \quad
    \begin{subfigure}[c]{0.22\textwidth}
        \includegraphics[width=\textwidth]{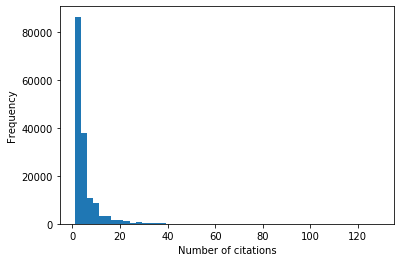}
        \subcaption{Average citation number.}
        \label{subfig:avg_no_citations}
    \end{subfigure}
    \caption{Characteristics of abstracts and citations in the datasets.}
    \label{fig:statistics_abstracts_citations}
\end{figure}

The final dataset contains only papers with an English abstract and referencing conference proceedings also contained in SciGraph. For training, publications from the years 1975-2014 were used, whereas for validation and testing we utilised the papers published in 2015 and 2016, respectively. Table~\ref{tab:characteristics_datasets} shows basic statistics of the three datasets. In total, the recommender system was trained on 1,122 distinct conference series and 137,376 papers written by 164,103 authors. Fig. \ref{fig:statistics_abstracts_citations} shows the average length of abstracts in the data (130 words) and the mean number of 5 citations per publication towards other conference proceedings in SciGraph\footnote{While the total number of citations in each paper is higher, not all cited papers are contained in SciGraph. These numbers reflect the edges in SciGraph, i.e. a paper in SciGraph citing another paper also in SciGraph.}.

\begin{table*}[htbp]
\caption{Characteristics of the training, validation, and test sets.}
\label{tab:characteristics_datasets}
\centering
\resizebox{\textwidth}{!}{%
\renewcommand{\arraystretch}{1.2}
\begin{tabular}{|l|r|r|r|r|}
\hline
 & Training  (1975-2014) & Validation  (2015) & Test  (2016) & Overlap  (training - test) \\ \hline
 Distinct conference series IDs & 1,122 & 311 & 518 & 394 \\
Distinct author names & 164,103 & 19,994 & 30,175 & 14,639 \\ 
All papers & 293,836 & 130,002 & 213,000 & -- \\
Papers w/ English abstracts & 290,877 & 12,990 & 21,223 & -- \\
\begin{tabular}[c]{@{}l@{}}Papers w/ English abstracts \\ and citations in SciGraph\end{tabular} & 137,376 & 7,511 & 11,600 & --\\ 
\hline
\end{tabular}
}
\end{table*}

\subsection{WikiCfP and H5-Index}
\label{subsec:wikicfp_h5index}
Two more datasets were utilized only in the user interface, to augment the recommended conferences with additional information, such as conference dates or ranking. However, these datasets were not used for training the models or generating the recommendations, which was done solely on SciGraph.


One dataset consists of calls for papers (CfPs) published on WikiCfP\footnote{\url{http://www.wikicfp.com/cfp/} (last accessed 31/12/2019)}. We implemented a crawler to create a dataset of conference names, acronyms, dates, locations, submission deadlines (e.g. mandatory attributes), as well as the conference series, categorization in WikiCfP, textual description and links to the conference's webpage (e.g. optional attributes). Overall, 4,902 CfPs were crawled in January 2020, out of which nearly a third have missing mandatory information. The crawled data was linked to the SciGraph dataset using Damerau-Levenshtein string similarity between the conference names. For each matched SciGraph conference, we retrieved the latest corresponding CfP with a submission deadline after January~1\textsuperscript{st}, 2020. This resulted in 8.24\% SciGraph conferences being matched to WikiCfP. The low linking precision (0.086) and recall (0.09) are determined by the low number of CfPs that met this strict submission deadline condition and were published on WikiCfP at the time of crawling. 


The other dataset comprises of conference ratings published by Google Scholar Metrics\footnote{\url{https://scholar.google.de/citations?view_op=top_venues&hl=en} (last accessed 31/12/2019)}, and was utilised to provide users with information about the quality of conferences. As there is no downloadable version of the data, a web scraper was implemented to retrieve the h5-index of ranked conferences. Google Scholar Metrics offers these statistics in the form of fixed-sized tables with 20 rankings, for 8 categories and 194 subcategories. A dataset containing 4,540 publications' ratings was extracted. This was integrated in the SciGraph dataset using Damerau-Levenshtein string similarity on conference names, and resulted in 2.73\% of the SciGraph conferences to be enriched with a h5-index rating, with a precision of 0.34 and recall of 0.55. The small fraction of matches is caused by the low number of conferences included in the Google Scholar Metrics. Terms such as "Conference" or "Workshop" appear in only 3.9\% of the ratings, while the word "Journal" is explicitly mentioned in 32.8\% of the rated publications.

Since the matches were scarce, we decided to check them manually, and to provide the supplementary information from WikiCfP and Google Scholar Metrics in the user interface only for the correct matches.

\section{Recommendation Techniques}
\label{sec:recommendation_techniques}

We developed two classes of recommendation techniques. The baseline, an author-based model, creates content-based recommendations using the publication history of a paper's authors. 
The model aggregates, per conference series, all papers sharing at least one author name, and recommends unique conference series where the authors of a given paper have previously published\footnote{We do not distinguish authors by unique IDs here, since only a subset of all Springer Nature authors have been disambiguated.}. The results are ranked decreasingly by the count of publications per conference series. 

GNN-based models aim to project similar entities closely in a latent vector space based on their features and relationships to other entities. Thus, the models learn not only the topological structure of a local neighbourhood, but also the distribution of the node features among its neighbours. To this end, the input data is structured as a graph with the following characteristics: nodes represent conference proceedings, node features encode their textual content, node labels denote the conference where the paper was published, while the edges denote citations or co-authorship links. We constructed three graph variants based on the edge type. The \emph{citations graph} contains edges representing citations among connected papers. In the \emph{co-authorship graph}, an edge is incident in two nodes if the corresponding papers share at least two authors, if not specified otherwise. Lastly, the \emph{heterogeneous graph} incorporates both kinds of edges.  

\begin{figure*}[htbp]
\centering
\includegraphics[width=0.725\textwidth]{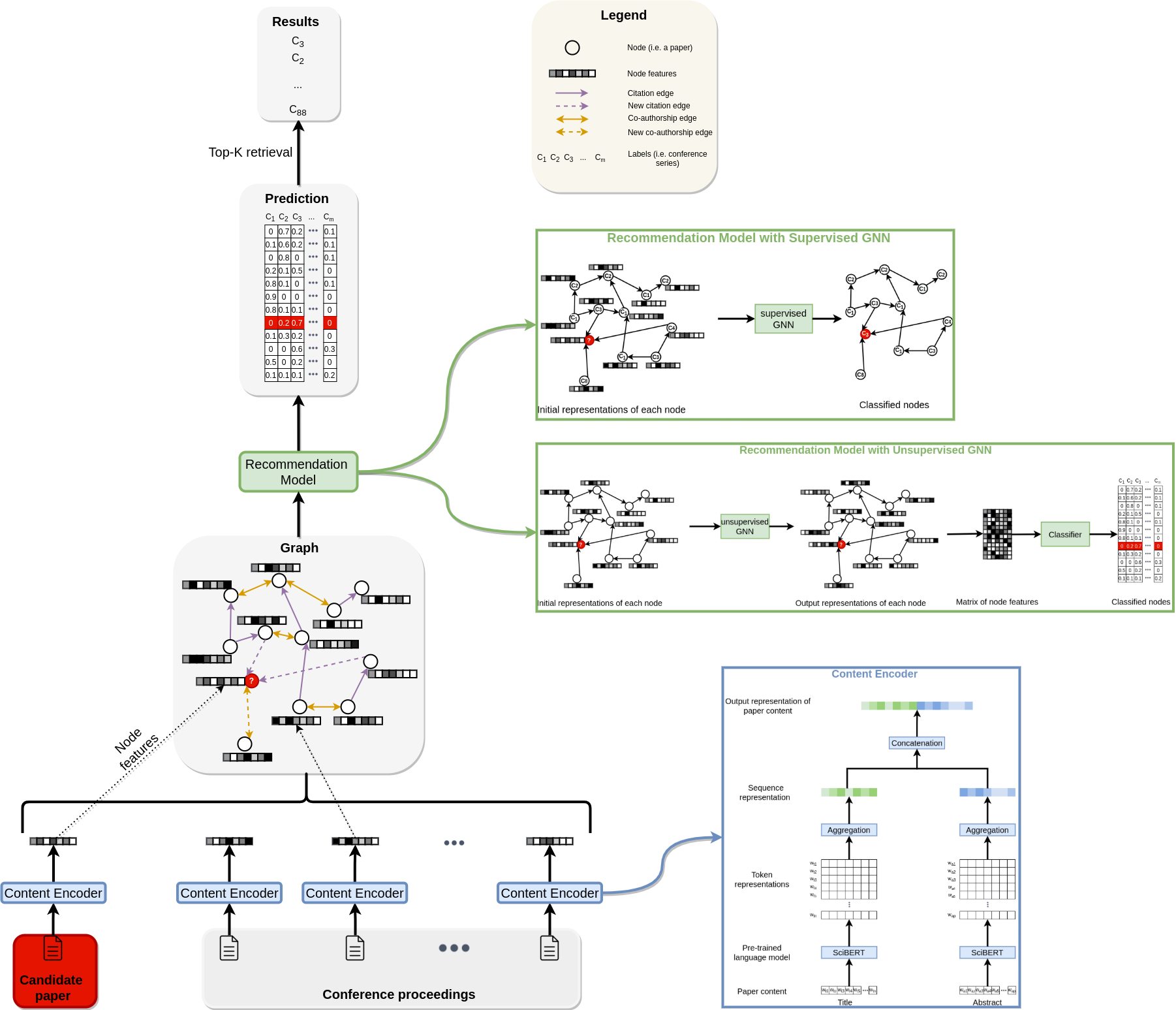}
\caption{Illustration of GraphConfRec's pipeline, using a \emph{heterogeneous graph.}}
\label{fig:pipeline}
\vspace{-0.5cm}
\end{figure*}

The recommendation problem is modelled as a \emph{node classification} task consisting in the prediction of a label for a new node denoting the input paper. More specifically, a prospective publication represents a new node that is attached to the existing graph by means of citation or co-authorship links. All models are inductive, meaning that they can handle unseen nodes, and either output node embeddings (unsupervised model), or directly compute a probability distribution over the classes (supervised model), utilizing the feature information and position of the new node in the graph. In the unsupervised approaches, a classifier is trained on the embeddings generated by the GNNs and utilised to classify the nodes. In all scenarios, the system recommends the top $N$ most probable conferences for a given paper. We implemented several versions of each recommendation technique based on the training method and graph type. Fig. \ref{fig:pipeline} illustrates the GraphConfRec's pipeline.

In earlier works \cite{iana_building_2019}, we observed that word embeddings trained on a scientific corpus perform better on SciGraph than those trained on general purpose corpora, such as Wikipedia, as the dataset mostly focuses on computer science publications. Therefore, we embed the papers' titles and abstracts using the pre-trained transformer-based \cite{vaswani_attention_2017} language model SciBERT \cite{beltagy_scibert_2019}, with an uncased vocabulary. 
Since SciBERT allows sequences of maximum 512 tokens, we assume that the most relevant information is contained within the first 512 words of an abstract or title, and the remaining words are ignored. This observation is consistent with the data, since only 31 paper abstracts from the training and validation sets are longer than 512 tokens. For every token, SciBERT outputs a vector of size 768, for each of its 12 hidden layers. Table \ref{tab:scibert_embeddings} enumerates the strategies utilised to compute a sequence's representation from its individual tokens. The resulting embeddings of titles and abstracts are then concatenated and treated as node features.

\subsubsection{GraphSAGE (SAmple and aggreGatE)-based models} compute a node’s embedding by aggregating the feature information of the nodes in its uniformly sampled, fixed-sized neighbourhood, using several learned aggregation functions proposed by Hamilton et al. \cite{hamilton_inductive_2017}. Five architectures are used: (1) mean aggregator -- computes the element-wise mean of the vectors; (2) mean and (3) max pooling aggregators -- apply an element-wise mean or max pooling operation on every neighbour's vector, after each of them have been passed through a fully-connected single-layer neural network; (4) a modified, inductive GCN aggregator \cite{kipf_semi-supervised_2016}; and (5) an LSTM-based aggregator \cite{hochreiter_long_1997}. We implemented different GraphSAGE-based recommendation models to examine the importance of training type and similarity computation on recommendations. \emph{GraphSAGE Neighbour} computes the papers’ similarity in the latent vector space and recommends conferences corresponding to the input paper’s nearest neighbours. \emph{GraphSAGE Classifier} utilises a classifier trained on the embeddings generated by the unsupervised GraphSAGE to predict conferences. We applied \emph{GraphSAGE Classifier} on both the \textit{citations graph} and the \textit{co-authorship graph} to analyse the influence of graph type on recommendations. Additionally, we implemented \emph{GraphSAGE Classifier Concat}, which concatenates the embeddings outputted by GraphSAGE trained on the citations and the co-authorship graphs, before feeding the resulting vector into a classifier. Lastly, \emph{GraphSAGE supervised} recommenders directly compute a probability distribution over the classes.

\subsubsection{GraphSAGE\_RL-based models} sample neighbourhoods based on a non-linear regression function which learns the importance of different candidate neighbourhoods, and thus, are able to sample only the most significant neighbours for a particular node, in order to avoid the high variance in training and inference, which can be caused by unbiasedly sampling using a uniform distribution \cite{oh_advancing_2019}. A reinforcement learning (RL) policy is used to determine the weight of each combination of node and neighbourhood, from the negative classification loss output of a GraphSAGE model.  We created several recommenders using Oh et al.'s \cite{oh_advancing_2019} GNN architecture, namely: \emph{GraphSAGE\_RL Classifier (citations graph)}, \emph{GraphSAGE\_RL supervised (citations graph)}, \emph{GraphSAGE\_RL supervised (heterogeneous graph)}.

\subsubsection{GAT (Graph attention network)-based models} compute a node’s latent representation by applying a self-attention mechanism over the features of its neighbours, which receive different weights based on their contribution, as proposed in \cite{velickovic_graph_2017}. We experimented with two variants of the GAT-based recommenders, namely \emph{GAT (citations graph)}, and \emph{GAT (heterogeneous graph)}, to obtain a probability distribution over the labels and to determine whether using the attention mechanism improves computational costs and recommendations's quality.

\subsubsection{HAN (Heterogeneous graph attention network)-based models} operate, in contrast to the approaches discussed above, on heterogeneous graphs, and explicitly distinguish between different edge types. Node-level attention weights meta-path-based neighbours based on their influence on a given node, whereas semantic-level attention distinguishes between different meta-paths given their importance. In Wang et al.'s \cite{wang_heterogeneous_2019} GNN architecture, the embedding of a node is computed by hierarchically aggregating, for all semantic-specific embeddings, the feature information of the meta-path neighbours. A HAN-based recommendation model was implemented to analyse whether explicitly modelling various edge types and meta-paths (e.g. paper-author-paper, paper-citation-paper) in the GNN improves the recommender's performance.

\subsubsection{SciBERT-ARGA-based models} adopt an approach inspired by Jeong et al. \cite{jeong_context-aware_2019}, and place a higher weight on the paper’s abstracts. The abstracts are embedded using SciBERT, as before, while the nodes of the citation or co-authorship networks are embedded using the adversarially regularised graph autoencoder (ARGA) framework proposed by Pan et al. \cite{pan_adversarially_2018}. The resulting vectors are concatenated and passed through a feed-forward neural network (FFNN), before a softmax function computes the probability distribution over conference labels. In ARGA, firstly a graph convolutional autoencoder reconstructs the topological graph structure from the embeddings generated from the node feature information and original graph structure \cite{pan_adversarially_2018}. ARGVA follows the same model architecture as ARGA, but the convolutional graph autoencoder (GAE) is replaced with a variational graph autoencoder (VGAE). Secondly, an adversarial training scheme is employed to enforce the latent representations generated by the GAE to match a prior Gaussian distribution \cite{pan_adversarially_2018}. More specifically, a generator and a discriminator play a minimax game in which the generator learns to produce “fake samples” which are as close as possible to the real ones, while the discriminator tries to distinguish whether an example comes from the true data distribution or is produced by the generator \cite{goodfellow_generative_2014}. Different variants of this recommender model are developed, using both the \emph{citations graph} and the \emph{heterogeneous graph}.

\section{Evaluation}
\label{sec:evaluation}

GraphConfRec's\footnote{Code available at \url{https://github.com/andreeaiana/graph_confrec}} performance was evaluated in two sets of experiments: a quantitative study on SciGraph, as well as a qualitative user study. 

\subsection{Experimental Setup}
\label{sec:experimental_setup}
The quantitative experiments were performed on a test set comprising of publication data from 2016. On the one hand, we analyse whether the proposed methods result in suitable conference recommendations for a prospective manuscript, as well as the results' ranking. For each publication in the test set, every model aims to predict the conference at which it has been published, and the results are compared against the truth value (i.e. the conference where it was published). For every technique, we generate 10 recommendations\footnote{Authors-based recommendations may create shorter lists if all authors have published at less than 10 conferences contained in SciGraph.}, and report the recall@10 and mean average precision (MAP) at different positions. Moreover, we experiment with different combinations of aggregating SciBERT embeddings and a wide range of model-specific parameter settings for each of the GNN-based recommendation techniques to determine the influence of each model component and hyperparameter setting on the overall results. On the other hand, we conducted a user study using an online prototype service to examine the quality of the recommendations from the researchers' perspective.


\subsection{Results of Quantitative Experiments}
\label{subsec:quantitative_results}

\input{scibert_and_results}

Table \ref{tab:best_results} summarizes the best results obtained with each model, as well as their configurations. Author-based recommendations perform reasonably well in terms of MAP, but have a low recall, as fewer than half of the recommendations contain the correct conference in the top 10 results. Two factors influence the results. Firstly, the number of recommendations is arbitrary and empty results lists are returned when none of the paper's authors overlap with those of the publications from the training set. This happens in 12.63\% of the cases due to the training-test sets split and illustrates the cold-start problem of the model, namely that users without publications in SciGraph cannot receive a recommendation. Secondly, the model cannot generate accurate predictions if none of the authors have published at a conference before, or if they only participated in events taking place after 2014. 

In comparison, GNN-based recommendations are more diverse, considering not only the paper’s textual content, as summarised by its title and abstract, but also the relationship with other conference proceedings, cited works, or co-authored publications. However, not all GNN-based models outperform the authors-based recommendations. As aforementioned, the latter model may create shorter lists or be unable to generate any suggestions. In contrast, all other recommendation techniques compute recommendation lists of fixed length of 10 for all inputs, even if the author has no prior publication history. Overall, we observed that for the majority of the methods, SciBERT embeddings averaging or summing either the last, or the last two hidden layers, achieved the best results. This shows that generally, these final layers of the transformer suffice to summarise and encode the meaning of a sequence. 

Moreover, all GNN-based models, with the exception of the \emph{GraphSAGE Classifier (co-authorship graph)} outperform the second baseline, namely the \emph{GraphSAGE Neighbour} model, in terms of both recall and MAP. We explored different aggregation functions and classifiers for the GraphSAGE and GraphSAGE\_RL-based recommenders, namely K-Nearest Neighbours (KNN), Gaussian Naive Bayes (GNB), Multilayer Perceptron (MLP), and Multinomial Logistic Regression (MLR), as well as two reward policies for the reinforcement learning approach of the \emph{GraphSAGE\_RL supervised} models. The GCN aggregator generally achieved the best results for the GraphSAGE-based models, while the mean-concat aggregator performed better for the GraphSAGE\_RL-based recommenders. Among the classifiers, MLR and MLP achieve the best results, followed closely by KNN, with a faster training time, in exchange for a slight drop in performance. The \emph{GraphSAGE supervised} models were trained for 20 epochs each, while the other models were trained for 10 epochs, and the weights of the model with the lowest validation loss used for inference. All models from these two approaches sample 25 first-order and 10 second-order neighbours, and use hidden layers of dimensions 512. We noticed that sampling third-order neighbours or using smaller hidden dimensions decrease the performance.

In contrast to GraphSAGE-based methods using the \emph{citations graph}, the model trained solely on the co-authorship network yields a low performance, potentially determined by the graph's structure. In this case, the \emph{co-authorship graph} was built by connecting papers which share at least one author. This resulted in twice the number of connected components compared to the \emph{citations graph}, roughly 5 times more edges and an average node degree of 20 (versus 4 for the \emph{citations graph}). The large number of neighbours per node is prone to inducing noise in the model, since it is likely that not all papers written by an author share the same topic. Moreover, the large connected components act as clusters, meaning that conferences corresponding to papers outside of these groups are less likely to be recommended, even if they are relevant. Although the \emph{GraphSAGE supervised}-based techniques show a small increase in both recall and MAP, only the \emph{GraphSAGE\_RL supervised}-based recommenders outperform the \emph{Authors} model in terms of recall. Therefore, non-linearly sampling a node's neighbours is better suited for the citations networks, where not all the referenced papers have an equal influence on the work citing them.

In comparison, GAT-based recommendation techniques outperform the baselines in terms of both metrics. For these models, we experimented with different numbers of hidden units, attention heads, several dropout values, as well as residual connections. The models were trained using an early stopping strategy after 100 epochs with no change in the loss. Adopting the inductive parameter settings proposed in \cite{hamilton_inductive_2017} (i.e. two hidden layers with 256 units each, 4 heads per layer, for a total of 1024 features, followed by a layer with a single attention head which computes 1171 features, no dropout or residual connections) resulted in a lower performance than that of \emph{GraphSAGE Classifier (co-authorship graph)}. However, GraphConfRec's best performance (recall@10 of 0.580 and MAP@10 of 0.336) was obtained with a GAT model with only one hidden layer with 8 units and 8 heads, for a total of 64 features, dropout of 0.5, and residual connections, trained on a directed \emph{heterogeneous graph}. Increasing or decreasing the number of heads, dropout value, training only on the \emph{citations graph} or an undirected graph lowered both recall and MAP. The findings show that adding residual connections alleviates the vanishing gradients problem which might have caused the sudden drop in performance for the first configuration of the model. Moreover, a directed graph better encodes the structure of a citation network, which is naturally directed.

The heterogeneous graph attention model\footnote{We experimented with the same parameter settings as for the GAT model.} does not improve the results further, meaning that, in this scenario, treating edges differently based on their role (i.e. citation or co-authorship link) does not constitute a stronger signal for detecting related papers. Similarly, combining abstract and graph embeddings in the \emph{SciBERT-ARGA} model\footnote{We experimented with the number of hidden units in ARGA and FFNN.} could not outperform the GAT-based recommender. In turn, this proves that encoding the papers' titles and abstracts as node features is enough to represents a paper's textual content, and that placing a higher weight on the abstracts does not provide additional significant information for determining the most similar publications to a given paper. 

Another important factor to consider are the computational costs associated with training each class of recommenders. As illustrated in Table \ref{tab:best_results}, all recommendation models using an unsupervised GNN incur a long average training time, varying between 83 and 200 minutes per epoch. In contrast, methods with supervised GNNs are orders of magnitude faster, with an average training time ranging from 6 to 226 seconds per epoch. Moreover, we observed that models using the \emph{citations graph} are 8.7 (for \emph{GraphSAGE supervised}-based techniques) to 36 times (for \emph{GAT}-based models) faster during training than those utilizing the \emph{heterogeneous graph}. This is determined by the more complex structure and larger size of the \emph{heterogeneous graph}, which is characterised by a greater number of edges for the same number of nodes as the \emph{citations graph}.

\subsection{Results of User Study}
\label{subsec:qualitative_results}

The user study\footnotemark was conducted with a group of 25 academic researchers. More than 80\% of the participants were either PhD or Master students, with at least one publication, while the remaining subjects were post-docs or professors. Moreover, 16\% of the participants were contained in our graph, while the others had no publications contained in SciGraph. The respondents were asked to test four models (i.e. \emph{Authors}, \emph{GAT (citations graph)}, \emph{GAT (heterogeneous graph)}, \emph{HAN}) with papers of their choice, and rate the suitability of the recommendations on a scale from 1 (worst) to 5 (best), as well as to, optionally, provide comments about the quality of the suggestions. All participants sought recommendations for their own papers. 

\footnotetext{The study description was distributed in the Data and Web Science Group at the University of Mannheim, as well as on LinkedIn and Twitter.}


\begin{table}[t]
    \caption{Results of the user study.}
    \label{tab:user_study}
    \centering
    \begin{tabular}{|r|r|r|r|r|}
        \hline
         Model &  Authors & HAN & GAT (cit.) & GAT (het.) \\
         \hline
         Raters & 16 & 21 & 25 & 22 \\
         Avg. rating & 2.75 & \textbf{3.76} & 3.68 & 3.55 \\ \hline
    \end{tabular}
\end{table}

The results are shown in table~\ref{tab:user_study}. One important observation is that not all respondents provided feedback for all four models. This was most prevalent for the \emph{Authors} model, since authors without a publication history of their own or their co-authors contained in the graph, were given no recommendations by the baseline recommender. 

Generally, the participants considered the recommendations to be suitable, but that the ranking of the results could be improved, since suggestions perceived as relevant were often positioned lower in the list than those perceived as irrelevant. One positive aspect noticed by some of the respondents is that the \emph{GAT (citations graph)} and \emph{HAN} models were able to predict conferences not known by the authors, but which were highly suitable for the input paper's content, although no obvious clues were included in the abstract. Moreover, these models predicted fitting academic venues based sometimes solely on the paper's abstract and title, since the authors and citations were either not provided, or not found in the dataset. Additionally, including co-authorship links in the predictions did not consistently improve results, with some users ranking the citations-based GAT model higher. Moreover, the results of the two GAT-based models were usually nearly identical, with differences being noticed only in the ranking of the recommendations.

Furthermore, according to the users' feedback, the confidence scores displayed by the models were sometimes quite low, although the recommendations were of good quality. This indicates a mismatch between the probability scores yielded by the models, on the one hand, and the ratings from the user study, on the other hand. More specifically, recommendations rated positively in the user study still got fairly low confidence scores internally.

Another aspect noticed by a few of the participants is the strong focus on semantic-web-related conferences (e.g such conferences were recommended even if the topic of the paper was not related to this field), suggesting a bias towards this domain in the training data. The lack of conferences from different publishers, determined by the publisher-dependent dataset used, was also among the weaknesses mentioned by some of the users. Lastly, even though the models seem helpful in discovering interesting, less-known, or niche academic venues which are thematically-fitting, some users would prefer excluding suggestions concerning events organised either only once (e.g. workshops), or locally (i.e. only in a certain geographic region). This is determined by the researchers' desire to not only be aware of more well-known academic venues, but to also submit their work to prestigious conferences.

\subsection{Discussion of Results}
\label{subsec:discussion_results}

Even though the performance metrics reported in Table \ref{tab:best_results} do not exceed a recall of 0.580 and a MAP@10 of 0.336, these figures need to be analysed in a broader context. Overall, 23.9\% of the conference series from the test set do not overlap with those in the training data, meaning that for these there are no training signals. Since training features are generated only based on previous publications, this implies that all the approaches discussed in this paper cannot recommend conferences not contained in the training set. Therefore, the maximum achievable recall and MAP@10 values are 0.761.

Overall, we found that co-authorship relations constitute a meaningful signal for recommendations, and that a heterogeneous directed graph constitutes the best training dataset for the GNN-based models. Additionally, we observed that an attention mechanism which weighs neighbours differently based on their contributions to the central node (i.e. input paper) outperforms other GNN-based approaches at sampling neighbourhoods. Recommendations based on a graph attention network trained on a heterogeneous graph achieve a recall@10 of 0.580 and a MAP@10 of 0.336, while the other models are clearly inferior, obtaining at most a 9\% lower recall and MAP, using a reinforcement learning policy for training a node sampler, and a GCN-based aggregation function. 

One major caveat of the quantitative evaluation is that, by design, only one recommendation is counted as correct, since a paper can only be published at one conference. Nonetheless, multiple suggestions can be suitable (e.g. different machine learning conferences). 
Moreover, a paper might have been rejected by a conference which is considered relevant by GraphConfRec.
However, in the current setup, these would be regarded as false predictions, and precision is underestimated. Nevertheless, the results of the user study show that GraphConfRec infers meaningful suggestions, as most respondents assessed the recommendations to be of good or very good quality. In contrast to the results of the quantitative experiments, which indicate that the \emph{GAT (heterogeneous graph)} recommender achieves the highest performance in terms of both metrics, the respondents of the user study rated the \emph{HAN} model higher in the majority of cases. Moreover, they also appear to consider the recommendations generated by \emph{GAT (citations graphs)} to be slightly more suitable than those of the GAT model which incorporates authors in its predictions. The feedback of the user study shows that, while the HAN-based recommender achieves a lower performance on the test set, in a real-life scenario, it offers more varied, suitable, and often less-known recommendations than the GAT-based models.

\section{Limitations and Future Work}
\label{sec:limitations_future}

Although the experiments show promising results, GraphConfRec suffers from several limitations, for which we propose potential solutions in the following section. 

\subsection{Dataset and Publisher Independent Recommendations}
GraphConfRec recommends only conferences contained in SciGraph, thus excluding, among others, proceedings published at ACM, IEEE, AAAI, or ACL conferences. In turn, this assumes that authors firstly choose a publisher, which is often not the case in practice, as indicated by some of the user study respondents. Using a publisher-independent dataset, such as the Microsoft Academic Knowledge Graph \cite{farber_microsoft_2019}, could overcome this weakness. Another solution would be to use SciGraph only for training the recommendation models, and to recommend CfPs from a dataset like WikiCfP, instead of using the latter only for augmenting the conferences with external data. While this would allow the model to recommend new conferences that are not present in the training dataset, and cannot currently be suggested, it would lead to a new set of challenges, for example with regards to interlinking the datasets.

\subsection{Refining Confidence Scores and Rankings}
Some user study participants considered that the model's confidence scores were quite low, even when the suggestions were accurate. One solution would be to change the presentation of the scores, e.g. by re-scaling the confidences of the top 10 recommendations. 
Furthermore, some users have suggested that the ranking of results could be improved, with more suitable recommendations being placed higher in the list. This could be achieved by re-ranking only the recommended results. For example, the results could be clustered based on keywords describing conferences (e.g. extracted from CfPs categorizations), and the confidences of recommendations added per cluster. Thus, the conferences belonging to the cluster with the highest overall probability could be recommended before the ones belonging to clusters with lower confidences. The evaluation of the recommendation models shows that in most cases, the correct conference is recommended among the first 3 results. Moreover, for some topics, there might be less than 10 fitting conferences. Therefore, under this approach, the model will learn to cluster and recommend first the conferences most similar to the correct one. 

\subsection{Fine-tuning Recommendations based on Additional Features}
Another limitation of GraphConfRec regards the quality of external information provided in the user interface. Continuously crawling WikiCfP 
for information about future conferences could ensure that users are always provided with up-to-date information about the recommended conferences. Moreover, providing more comprehensive ratings of all the suggested conferences and incorporating different rankings, such as CORE\footnote{\url{http://portal.core.edu.au/conf-ranks/} (last accessed 23/04/2020)}, could provide an objective measure of distinguishing between numerous conferences from different tiers. However, removing niche conferences or workshops from the recommendation list could reduce the diversity of results. Overall, such external information would cater the individual needs of researchers by offering the option of fine-tuning the results based on various features, including conference ranking, submission deadline, or geo-based filtering.

\subsection{Exploring Other Graph Neural Networks}
Exploring other GNNs which handle heterogeneous graphs could represent another direction for future research. In contrast to HAN, which uses meta-paths manually defined by the user, the Graph Transformer Network proposed by Yun et al. \cite{yun_graph_2019}, can yield node representations without domain-specific graph pre-processing required for constructing and learning meta-paths. A different GNN for heterogeneous graphs, which avoids the weaknesses of handcrafted meta-paths, was developed by Hong et al. \cite{hong_attention-based_2020}. Their model automatically extracts information from heterogeneous graphs and learns meta-paths, by firstly projecting the transformations between nodes of different types into a low-dimensional entity space, and then, by aggregating this multi-relational information of the projected neighbourhood using a GNN \cite{hong_attention-based_2020}. 


\section{Conclusion}
\label{sec:conclusion}


In this paper, we introduced GraphConfRec, a recommender system for conferences, based on both a paper's textual content, and its co-authorship and citation networks. The system is intended for researchers searching for conferences to publish at. The recommendations are computed using SciGraph, with submission deadlines added from WikiCfP and conference rankings from Google Scholar Metrics. We found that a graph attention network-based recommender trained on a heterogeneous graph achieves the best performance, with a recall@10 of 0.580 and a MAP@10 of 0.336. The user study showed that GraphConfRec infers meaningful recommendations, its GNN-based predictions achieving an overall average score of 3.7 out of 5. However, users consider recommendations based on the heterogeneous graph attention network model to be more diverse, and often, slightly more thematically-fitting for their papers than those of the GAT-based recommender.


Although abstract, title, citations and co-authorships provide useful signals for recommendation, other entities, such as keywords or affiliations, could be incorporated to augment the data. One of the weaknesses of the current approach is considering only referenced conference proceedings found in SciGraph. Including information about cited books and journals, both from SciGraph and other publishers, could improve the models' performance to a certain extent. Furthermore, a publisher-independent dataset could further improve GraphConfRec's performance and extent its usability to a larger group of researchers which seek to publish in conference proceedings from other institutions. Lastly, improving the ranking of the results, providing fine-tuning of recommendations based on conference features, or exploring other GNN models constitute potential avenues for future research.

In conclusion, while additional options should be explored in future work, the results offer a first glance at how semantic data and graph neural networks can be utilised to build a recommender system for conferences. Hence, it opens promising future directions, such as extending the system into a full-fledged recommendation engine for scientific publications.

\bibliographystyle{ieeetr}
\bibliography{references}

\end{document}

%% file: scibert_and_results.tex
\begin{sidewaystable*}
\caption{Approaches for computing the SciBERT embeddings of the papers.}
\label{tab:scibert_embeddings}
\centering
\scriptsize
\resizebox{0.6\textwidth}{!}{%
\renewcommand{\arraystretch}{1.2}
\begin{tabular}{|l|l|}
\hline
 Notation & Description \\ \hline
 AVG\_L & Mean of last hidden layer representation of all tokens \\
AVG\_2L & Mean of second-to-last hidden layer representation of all tokens \\ 
AVG\_SUM\_L4 & Mean of the sum of the last four hidden layers representation of all tokens \\ 
AVG\_SUM\_ALL & Mean of the sum of all the hidden layers representation of all tokens \\ 
MAX\_2L & Max pool over the second-to-last hidden layer representation of all tokens \\ 
CONC\_AVG\_MAX\_2L & Concatenation of the AVG\_2L and MAX\_2L vectors \\ 
CONC\_AVG\_MAX\_SUM\_L4 & Concatenation of the AVG\_SUM\_L4 and MAX\_SUM\_L4 vectors \\ 
SUM\_L & Sum of the last hidden layer representation of all the tokens \\ 
SUM\_2L & Sum of the second-to-last hidden layer representation of all the tokens \\
\hline
\end{tabular}%
}
\\

\vspace*{1 cm}

\caption{Results of the best performing individual recommendation techniques.
For each model, we only report the results of the best performing model configuration.}
\label{tab:best_results}
\scriptsize
\resizebox{\textwidth}{!}{%
\renewcommand{\arraystretch}{1.2}
\begin{tabular}{|l|l|l|c|c|c|c|c|c|}
\hline
 & \multicolumn{2}{|c|}{Model configuration} & \multicolumn{2}{|c|}{Training statistics} & \multicolumn{4}{|c|}{Evaluation results}\\ \cline{2-3} \cline{4-5}  \cline{6-9}
Model & Embedding type & Parameter settings & Avg. time (s) per epoch & Epochs trained & Recall@10 & MAP@10 & MAP@5 & MAP@3 \\ \hline
Authors (1\textsuperscript{st} baseline) & -- & -- & -- & -- & 0.458 & 0.308 & 0.302 & 0.290 \\ \hline
GraphSAGE Neighbour (2\textsuperscript{nd} baseline) & CONC\_AVG\_MAX\_SUM\_L4 & maxpool aggregator & 11,994.46 & 10 & 0.259 & 0.096 & 0.082 & 0.072 \\ 
GraphSAGE Classifier (citations graph) & CONC\_AVG\_MAX\_2L & GCN aggregator + MLR & 4,991.29 & 10 & 0.414 & 0.244 & 0.234 & 0.221 \\ 
GraphSAGE Classifier (co-authorship graph) & AVG\_SUM\_ALL & maxpool aggregator + MLR & 9,010.57 & 10 & 0.054 & 0.019 & 0.015 & 0.014 \\ 
GraphSAGE Classifier Concat & SUM\_L & GCN aggregator + KNN (n=30) & 3,271.12 & 10 & 0.395 & 0.237 & 0.228 & 0.215 \\ 
GraphSAGE supervised (citations graph) & AVG\_2L & GCN aggregator & 29.63 & 20  & 0.417 & 0.246 & 0.236 & 0.223  \\ 
GraphSAGE supervised (heterogeneous graph)  & AVG\_SUM\_ALL & GCN aggregator & 32.48 & 20 & 0.440 & 0.258 & 0.247 & 0.234\\ \hline
GraphSAGE\_RL Classifier (citations graph) & SUM\_L & GCN aggregator + MLR & 10,699.81 & 10 & 0.414 & 0.242 & 0.231 & 0.220 \\ 
GraphSAGE\_RL supervised (citations graph) & AVG\_L & mean-concat aggregator + last-hop reward & 36.43 & 10 & 0.531 & 0.298 & 0.284 & 0.266  \\ 
GraphSAGE\_RL supervised (heterogeneous graph) & SUM\_L & mean-concat aggregator + all-hops reward & 56.49 & 10 & 0.546 & 0.306 & 0.292 & 0.273  \\ \hline
GAT (citations graph) & AVG\_L & 8 attention heads with 64 hidden units each & 93.53 & 367 & 0.572 & 0.327 & 0.312 & 0.295 \\ 
\textbf{GAT (heterogeneous graph)} & SUM\_2L & 8 attention heads with 64 hidden units each & 147.25 & 503 & \textbf{0.580} & \textbf{0.336} & \textbf{0.322} & \textbf{0.303} \\ \hline
HAN & AVG\_L & 8 attention heads with 128 hidden units each & 226.34 & 301 & 0.540 & 0.300 & 0.285 & 0.267 \\ \hline
SciBERT-ARGA (citations graph) & AVG\_2L & ARGVA + FFNN with 500 hidden units & 5.71 & 200 & 0.530 & 0.293 & 0.278 & 0.261 \\ 
SciBERT-ARGA (heterogeneous graph) & AVG\_L & ARGA + FFNN with 500 hidden units & 6.57 & 200 & 0.534 & 0.295 & 0.280 & 0.263  \\ \hline
\end{tabular}%
}
\end{sidewaystable*}